\documentclass[a4paper]{jpconf}
\usepackage{graphicx}
\begin{document}
\title{Implication of Higgs mediated Flavour Changing Neutral Currents with Minimal Flavour Violation}

\author{M. N. Rebelo}

\address{Centro de F\' \i sica Te\' orica de Part\' \i culas (CFTP),
Instituto Superior T\' ecnico (IST), U. de Lisboa (UL), Av. Rovisco Pais, P-1049-001 Lisboa,
Portugal.}

\ead{rebelo@tecnico.ulisboa.pt}

\begin{abstract}
We analise phenomenological implications of two Higgs doublet models with Higgs 
flavour changing neutral currents suppressed in the quark sector by small entries of 
the Cabibbo-Kokayashi-Maskawa matrix. This suppression occurs in a natural way
since it is the result of a symmetry applied to the Lagrangian. These type of models
were proposed some time ago by Branco Grimus and Lavoura. Our results clearly show
that these class of models allow for new physical scalars, with masses which are
reachable  at the LHC. The imposed symmetry severely reduces the number of free 
parameters and allows for predictions. Therefore these models can eventually be proved right or 
eliminated  experimentally. 
\end{abstract}

\section{Introduction}
There are several good motivations to consider models with two Higgs doublets. Such models allow 
for new sources of CP violation and for the possibility of having spontaneous CP violation. 
It is by now established that the SM cannot account for the observed baryon asymmetry of the universe
and that new sources of CP violation are required.  Spontaneous CP violation 
was first proposed by Lee \cite{Lee:1973iz}
in the context of two Higgs doublet models (2HDM) and  puts CP violation 
on the  same footing as the electroweak symmetry breaking.
Models with two Higgs doublets may also provide a solution to the strong CP problem of the type proposed by 
Peccei and Quinn \cite{Peccei:1977hh}. 
Furthermore, supersymmetric extensions of the Standard Model (SM) also require the 
existence of two Higgs doublets.  \\

The discovery of a scalar boson in the run 1 of LHC by Atlas \cite{Aad:2012tfa}  
and CMS  \cite{Chatrchyan:2012ufa}  immediately raises the question of whether this is the SM Higgs 
boson or part of a multi-Higgs theory. The properties of the Higgs boson already discovered are up till 
now in good agreement with those predicted by the SM but  more precise determinations may show evidence for
New Physics. There is also the possibility that the LHC will  soon discover a charged Higgs or additional
neutral scalars thus confirming the need to extend the scalar sector of the SM. \\

There are important experimental constraints on 2HDM. Such models have potentially large 
Higgs mediated flavour changing neutral currents (FCNC)  \cite{Gunion:1989we}, \cite{Branco:2011iw}
\cite{Crivellin:2013wna}.
Effects due to these FCNC are severely constrained and therefore some mechanism to suppress these effects 
is required. It is possible to eliminate tree level FCNC by imposing for instance natural flavour
conservation  \cite{Glashow:1976nt} or  alignment \cite{Pich:2009sp}.  Another possibility, which was
proposed some  time ago by Branco Grimus and Lavoura (BGL) \cite{Branco:1996bq},  
consists on imposing a symmetry on the Lagrangian allowing
for tree level FCNC in the quark sector suppressed by small entries of the 
Cabibbo-- Kobayashi -- Maskawa matrix, $V_{CKM}$.  Later-on, BGL models were
extended to the leptonic sector   \cite{Botella:2011ne} and their relation to Minimal Flavour Violation
models has been studied  \cite{Botella:2009pq}.  Phenomenological implications of these models
have been analysed recently \cite{Bhattacharyya:2013rya}, \cite{Botella:2014ska},  \cite{Bhattacharyya:2014nja}.
This talk is largely based on work done in Ref. \cite{Botella:2014ska}.

\section{Theoretical Framework}
In Ref.~\cite{Botella:2014ska} we
analysed extensions of the SM with two Higgs doublets together with three right-handed neutrinos.
We did not add Majorana mass terms to the Lagrangian and as a result neutrinos are Dirac type.
However our analysis of phenomenological implications is not sensitive to the
character of the neutrinos. The Yukawa  interactions can be explicitly written:
\begin{eqnarray}
{\mathcal{L}}_{Y} &=&-\overline{Q_{L}^{0}}\ \Gamma _{1}\Phi _{1}d_{R}^{0}-
\overline{Q_{L}^{0}}\ \Gamma _{2}\Phi _{2}d_{R}^{0}-\overline{Q_{L}^{0}}\
\Delta _{1}\tilde{\Phi }_{1}u_{R}^{0}-\overline{Q_{L}^{0}}\ \Delta _{2}
\tilde{\Phi }_{2}u_{R}^{0}  \nonumber \\
&&-\overline{L_{L}^{0}}\ \Pi _{1}\Phi _{1}\ell_{R}^{0}-\overline{L_{L}^{0}}\
\Pi _{2}\Phi _{2}\ell_{R}^{0}-\overline{L_{L}^{0}}\ \Sigma _{1}\tilde{\Phi} _{1}
\nu_{R}^{0}-\overline{L_{L}^{0}}\ \Sigma_{2}\tilde{\Phi}_{2}\nu_{R}^{0}+\mbox{h.c.}, 
\label{YukawaDirac1}
\end{eqnarray}
where $\Gamma_i$, $\Delta_i$  $ \Pi_{i}$ and $\Sigma_{i}$ denote the Yukawa couplings
to the right-handed quarks $d^0_R$, $u^0_R$, rigth-handed leptons $\ell_{R}^{0}$, 
$\nu_{R}^{0}$, respectively, and the Higgs doublets $\Phi_j$. The quark mass matrices generated
after spontaneous gauge symmetry breaking are given by:
\begin{eqnarray}
M_d = \frac{1}{\sqrt{2}} ( v_1  \Gamma_1 +
                           v_2  e^{i \alpha}   \Gamma_2 ), \quad 
M_u = \frac{1}{\sqrt{2}} ( v_1  \Delta_1 +
                           v_2  e^{-i \alpha} \Delta_2 ),
\label{mmmm}
\end{eqnarray}
where $v_i \equiv |<0|\phi^0_i|0>|$ and $\alpha$ denotes the relative phase 
of the vacuum expectation values (vevs) of the neutral components of 
$\Phi_i$. 
The matrices $M_d$,  $M_u$ are diagonalized by the usual bi-unitary transformations:
\begin{eqnarray}
U^\dagger_{dL} M_d U_{dR} = D_d \equiv {\mbox diag}\ (m_d, m_s, m_b) 
\label{umu}\\
U^\dagger_{uL} M_u U_{uR} = D_u \equiv {\mbox diag}\ (m_u, m_c, m_t)
\label{uct}
\end{eqnarray}
The neutral and the charged Higgs interactions obtained from Eq.~(\ref{YukawaDirac1}) 
for the quark sector are of the form
\begin{eqnarray}
{\mathcal L}_Y (\mbox{quark, Higgs})& = & - \overline{d_L^0} \frac{1}{v}\,
[M_d H^0 + N_d^0 R + i N_d^0 I]\, d_R^0  - \nonumber \\
&-& \overline{{u}_{L}^{0}} \frac{1}{v}\, [M_u H^0 + N_u^0 R + i N_u^0 I] \,
u_R^{0}  - \label{rep}\\
& - &  \frac{\sqrt{2} H^+}{v} (\overline{{u}_{L}^{0}} N_d^0  \,  d_R^0 
- \overline{{u}_{R}^{0}} {N_u^0}^\dagger \,    d_L^0 ) + \mbox{h.c.} \nonumber 
\end{eqnarray}
where $v \equiv \sqrt{v_1^2 + v_2^2} \approx \mbox{246 GeV}$,  
 and $H^0$, $R$ are orthogonal combinations of the fields  $\rho_j$,
arising when one expands  \cite{Lee:1973iz}  the neutral scalar fields
around their vacuum expectation values, $ \phi^0_j =  \frac{e^{i \alpha_j}}{\sqrt{2}} 
(v_j + \rho_j + i \eta_j)$, choosing $H^0$ in such a way
that it has couplings to the quarks which are proportional
to the mass matrices, as can be seen from Eq.~(\ref{rep}). 
Similarly, $I$ denotes the linear combination
of $\eta_{j}$ orthogonal to the neutral Goldstone boson. 
The matrices $N_d^0$,  $N_u^0$  
are given by:
\begin{eqnarray}
N_d^0 = \frac{1}{\sqrt{2}} ( v_2  \Gamma_1 -
                           v_1 e^{i \alpha} \Gamma_2 ), \quad 
N_u^0 = \frac{1}{\sqrt{2}} ( v_2  \Delta_1 -
                           v_1 e^{-i \alpha} \Delta_2 )
\end{eqnarray}
It is clear from these expressions that the  flavour structure of the quark sector of  two 
Higgs doublet models is much richer than that of the SM requiring the 
four matrices $M_d$,  $M_u$, $N_d^0$,  $N_u^0$ in order to be fully specified.
For the leptonic sector one can derive similar expressions and
the corresponding matrices can be denoted by $M_\ell$, $M_\nu$, $N_\ell^0$, $N_\nu^0$.
In the leptonic sector with Dirac neutrinos the analogy with the quark sector is perfect.

Flavour changing neutral currents in the quark sector  are controlled by $N^0_d$ and 
$N^0_u$ while in the leptonic sector they are controlled by $N_\ell^0$, $N_\nu^0$.

In terms of physical quarks the neutral and the charged Higgs interactions can be written:
\begin{eqnarray}
{\mathcal L}_Y (\mbox{quark, Higgs} )&  =  &  
 - \frac{\sqrt{2} H^+}{v} \bar{u} \left(
V N_d \gamma_R - N^\dagger_u \ V \gamma_L \right) d +  \mbox{h.c.} -  \nonumber  \\
& - & \frac{H^0}{v} \left(  \bar{u} D_u u + \bar{d} D_d \ d \right) -   
 \frac{R}{v} \left[\bar{u}(N_u \gamma_R + N^\dagger_u \gamma_L)u+
\bar{d}(N_d \gamma_R + N^\dagger_d \gamma_L)\ d \right] + \nonumber \\
& + &  i  \frac{I}{v}  \left[\bar{u}(N_u \gamma_R - N^\dagger_u \gamma_L)u-
\bar{d}(N_d \gamma_R - N^\dagger_d \gamma_L)\ d \right] 
\end{eqnarray}
where $\gamma_{L}$ and $\gamma_{R}$ are the left-handed and right-handed chirality projectors, 
respectively, and 	$N_d \equiv  U^\dagger_{dL} N_d^0 U_{dR}$, $N_u \equiv
U^\dagger_{uL} N_u^0 U_{uR}$, $V\equiv U^\dagger_{uL}U_{dL}$.  The matrix  $V$ 
is a simplified notation for the $V_{CKM}$ matrix.
There are analogous expressions for the leptonic sector with  $V_{CKM}$ replaced by the
Pontcorvo-Maki-Nakagawa-Sakata matrix, $U_{PMNS}$.
The physical neutral Higgs fields are combinations  of  $H^0$, $R$ and $I$. \\

Up till this point the discussion applies to the general two Higgs doublet model with Dirac fermions.
The matrices $N_d$, $N_u$, $N_\ell$ and $N_\nu$ are entirely arbitrary and the
scalar potential is the most general one for two Higgs doublets. \\

In order suppress the tree level FCNC in the quark sector by means of small entries of $V_{CKM}$
Branco, Grimus and Lavoura imposed the following symmetry on the quark and scalar sector of the
Lagrangian \cite{Branco:1996bq}:
\begin{equation}
Q_{Lj}^{0}\rightarrow \exp {(i\tau )}\ Q_{Lj}^{0}\ ,\qquad
u_{Rj}^{0}\rightarrow \exp {(i2\tau )}u_{Rj}^{0}\ ,\qquad \Phi
_{2}\rightarrow \exp {(i\tau )}\Phi_{2}\ ,  \label{S symetry up quarks}
\end{equation}
where $\tau \neq 0, \pi$, with all other quark fields transforming 
trivially under the symmetry. The index $j$ can be fixed as either 1,
2 or 3. Alternatively the symmetry may be chosen as:
\begin{equation}
Q_{Lj}^{0}\rightarrow \exp {(i\tau )}\ Q_{Lj}^{0}\ ,\qquad
d_{Rj}^{0}\rightarrow \exp {(i2\tau )}d_{Rj}^{0}\ ,\quad \Phi
_{2}\rightarrow \exp {(- i \tau)}\Phi_{2}\ .  \label{S symetry down quarks}
\end{equation} 
The symmetry given by Eq.~(\ref{S symetry up quarks}) leads to Higgs
FCNC in the down sector only, whereas the symmetry specified by 
Eq.~(\ref{S symetry down  quarks}) leads to Higgs FCNC only in the up
sector. These two alternative choices of symmetry combined with the
three possible ways of fixing  the index $j$ give rise to six
different realisations of 2HDM with the flavour structure, in the
quark sector, controlled by the $V_{CKM}$ matrix. The models obtained from
the symmetry defined by Eq.~({\ref{S symetry up quarks})  are called up-type models.
In these models $N_d$ and $N_u$ have the simple form:
\begin{equation}
(N_d)_{rs}= \frac{v_2}{v_1} (D_d)_{rs} - 
\left( \frac{v_2}{v_1} +  \frac{v_1}{v_2}\right) 
(V^\dagger_{CKM})_{rj} (V_{CKM})_{js} (D_d)_{ss} \label{24}
\end{equation}
no sum in $j$ implied, whereas, particularising the index $j$ to be 3, we have:
\begin{equation}
N_u  = - \frac{v_1}{v_2} \mbox{diag} \ (0, 0, m_t) +  \frac{v_2}{v_1}
\mbox{diag} \ (m_u, m_c, 0) \label{25} 
\end{equation}
the index $j$ fixes the row of $V_{CKM}$ which suppresses the flavour changing 
neutral currents.  For down-type models, which are those obtained from imposing 
the symmetry defined by Eq.~(\ref{S symetry down quarks}), the two matrices 
$N_d$ and  $N_u$  exchange r\^ ole and the FCNC are now suppressed by one of the
columns of  $V_{CKM}$ depending on the index $j$.

As a  result of imposing such a  symmetry the matrices $N_d$ and  $N_u$ are entirely 
determined by fermion masses, the $V_{CKM}$ matrix  
and the angle $\beta$ defined by $\tan \beta = {v_2}/{v_1}$, with no other free parameters. The flavour 
structure of BGL models depends on parameters already present in the SM
apart from the new parameter $\tan \beta$. This characteristic is a defining 
feature  of models that have been later-on denoted as models of Minimal Flavour Violation type 
\cite{Buras:2000dm}, \cite{D'Ambrosio:2002ex}, \cite{Bobeth:2005ck}, \cite{Dery:2013aba}.

The leptonic sector with Dirac neutrinos is analogous to the quark sector and again there are 
six possible different realisations. Combining the two sectors one obtains thirty six different
models which can be identified by a set of two indices.
For example, the model $(\mbox{up}_3, \ell_2) \equiv (t,\mu)$  will 
have no tree level neutral flavour changing couplings in the up quark and the charged lepton sectors 
while the neutral flavour changing couplings in the down quark and neutrino sectors will be controlled, 
respectively, by $V_{td_i}\ V_{td_j}^\ast$ and $U_{\mu \nu_a} \ U_{\mu \nu_b}^\ast$. \\

The scalar potential is also constrained by the imposed  symmetry. With the introduction of a soft
symmetry breaking term it will have seven independent parameters which will determine the four
scalar masses, the combination $v \equiv \sqrt{v_1^2 + v_2^2} $, $\tan \beta \equiv v_2/v_1$, and $\alpha$.
The angle $\alpha$ is the mixing angle relating the  physical neutral CP-even scalars to 
the fields $\rho_1$ and $\rho_2$. One mixing angle is sufficient since this
constrained scalar potential does not violate CP neither explicitly nor spontaneously \cite{Branco:1996bq}
and therefore the field $I$ is already physical.
The soft symmetry breaking term prevents the appearance of an would-be Goldstone boson 
due to an accidental continuous global symmetry of the potential. 

In the present work we assumed that $H^0$ coincides with the observed Higgs boson. Deviations from this
assumption are still allowed by the experimental data but are constrained to be small. This assumption
corresponds to imposing $\beta - \alpha = \pi / 2$. The masses of the extra Higgs bosons 
must obey constraints coming from electroweak precision tests, in particular the $T$ and $S$ 
parameters. Bounds on  $T$ and $S$  together with direct mass limits, significantly constrain the
masses of the new scalar fields in terms of  the mass of the charged Higgs $H^\pm$ 
\cite{Grimus:2008nb}    so that once this mass
is fixed there is not much freedom left for the masses of the extra neutral scalars. As a result
we can approximately scan the whole region of parameter space by varying $\tan \beta$ and 
and the mass, $m_{H^\pm}$, of the charged Higgs boson. 

\section{Confrontation with the experimental results}
We performed an analysis of the thirty six BGL models in order to determine where could the 
masses of the new scalars lie and how these depend on $\tan \beta$.  The masses of all new three 
scalar fields were treated independently and on an equal footing even though for simplicity
we only presented results in terms of $m_{H^\pm}$.  As stated in the previous section we assumed that
the discovered Higgs at the LHC coincides with our $H^0$ scalar, the one without FCNC.

We imposed present constraints from several relevant flavour observables. 
In Table 1 we summarise the different types of relevant observables
that we took into consideration, indicating where the contributions come from. In some 
cases the  new contributions are only present in some but not all of the models.
In Ref.~\cite{Botella:2014ska} we give a detailed description of the analysis that we performed
and we list the set of experimental data and bounds that we used.

\begin{table}[h] 
\begin{center}
\begin{tabular}{c|c|c|c|c|c|c|}
\cline{2-7} & \multicolumn{4}{|c|}{BGL - 2HDM} & \multicolumn{2}{|c|}{SM}\\ 
\cline{2-7} & \multicolumn{2}{|c|}{Charged $H^\pm$} & \multicolumn{2}{|c|}{Neutral $R$, $I$} &   &  \\
\cline{2-5} & \multicolumn{1}{|c|}{Tree} & \multicolumn{1}{|c|}{Loop} & \multicolumn{1}{|c|}{Tree} & \multicolumn{1}{|c|}{Loop} & \multicolumn{1}{|c|}{Tree} & \multicolumn{1}{|c|}{Loop} \\
\hline
\multicolumn{1}{|c|}
{$M\to\ell\bar\nu,M^\prime\ell\bar\nu$} &
 X &  $x$ &  &  $x$ &  X &  $x$ \\
\hline \multicolumn{1}{|c|} {Universality} & X 
& $x$ & 
&  $x$ & X& $x$ \\
\hline \multicolumn{1}{|c|} {$M^0\rightarrow \ell_1^+\ell_2^-$} &  &  $x$ & X & $ x$ &  & X \\
\hline  \multicolumn{1}{|c|}{$M^0\leftrightarrow \bar M^0$} & & $x$  & X & $x$ &   &  X \\
\hline  \multicolumn{1}{|c|} {$\ell_1^-\rightarrow \ell_2^-\ell_3^+\ell_4^-$} &  &  $x$  & X & $x$  &   & $ x$ \\
\hline \multicolumn{1}{|c|}{$B\rightarrow X_{s}\gamma$} &  & X &  &X &  & X\\
\hline \multicolumn{1}{|c|} {$\ell_j\rightarrow \ell_i\gamma$}  &   & X  &  &  X &  & $x$\\
\hline \multicolumn{1}{|c|}{ EW Precision} &   & X &  &  X & &  X \\
\hline
\end{tabular}
\end{center}
\caption{Summary table of the different types of relevant observables; leading contributions are tagged  X while subleading or negligible ones are tagged $x $.\label{TAB:summary}}
\end{table}

Figues 1 and 2 present the allowed regions  we obtained for each one of the thirty six models.

Some of the  BGL models allow for masses of the charged Higgs below 380 GeV which is 
the constraint from $b \rightarrow s \gamma$  on type II 2HDMs \cite{Hermann:2012fc}. 
This is due to the different dependence that these models have on $\tan \beta$ .
However, in general,  loop level processes, such as  $b \rightarrow s \gamma$  
as well as  $\ell_j \rightarrow \ell_i \gamma$ provide important constraints. The same is true for 
the  process $Z \rightarrow b \bar b$ and the oblique parameters $S$ and $T$,  unlike $U$.  

Concerning electric  dipole moments of leptons and quarks \cite{Raidal:2008jk}, \cite{Jung:2013hka}
BGL models do not give new physics one loop contributions.  In \cite{Botella:2012ab}
it has been shown that the weak basis invariant relevant for the quark EDMs 
does not acquire an imaginary part. In fact the same applies to two loop contributions
within BGL models..

\section{Conclusions}
BGL models are very constrained since they have a very small number of free parameters
and therefore they are highly predictive allowing to establish correlations among different observables. 
This also means that  in principle it will be possible to rule out several of these different scenarios
based on the future LHC results. Our results show that there are some very promising  BGL
implementations, which deserve more attention. 

Several of the models allow for scalar masses within the reach of direct searches  at the LHC.

\clearpage
\begin{figure}[!htb]
\centering
\includegraphics[width=\textwidth]{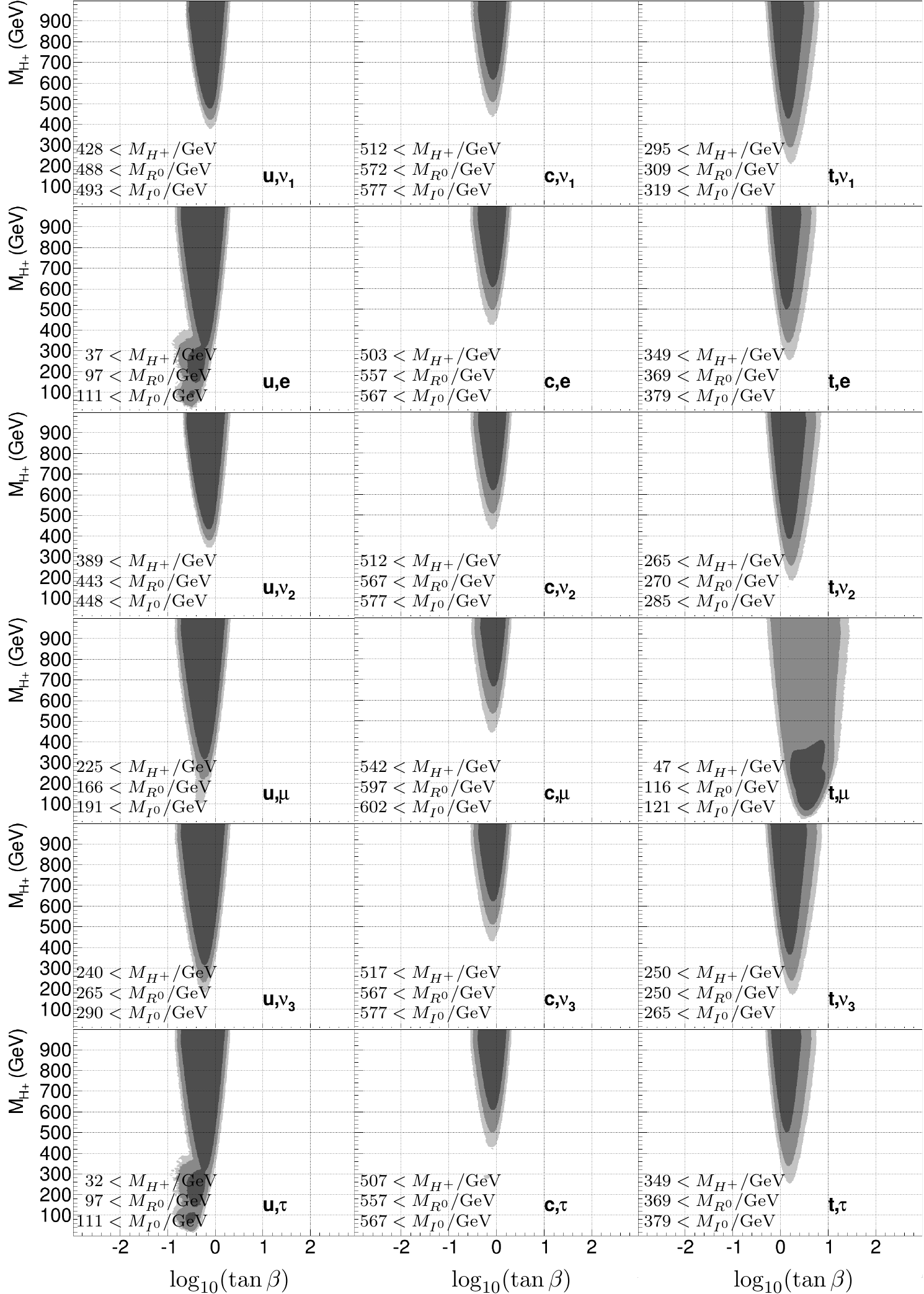}
\caption{Allowed 68\% (black), 95\% (gray) and 99\% (light gray) CL regions in $m_{H^\pm}$ vs. $\tan\beta$ for BGL models of types $(u_i,\nu_j)$ and $(u_i,\ell_j)$, i.e. for models with FCNC in the down quark sector and in the charged lepton or neutrino sector (respectively). Lower mass values corresponding to 95\% CL regions are shown in each case.\label{FIG:Results01}}
\end{figure}

\clearpage
\begin{figure}[!htb]
\centering
\includegraphics[width=\textwidth]{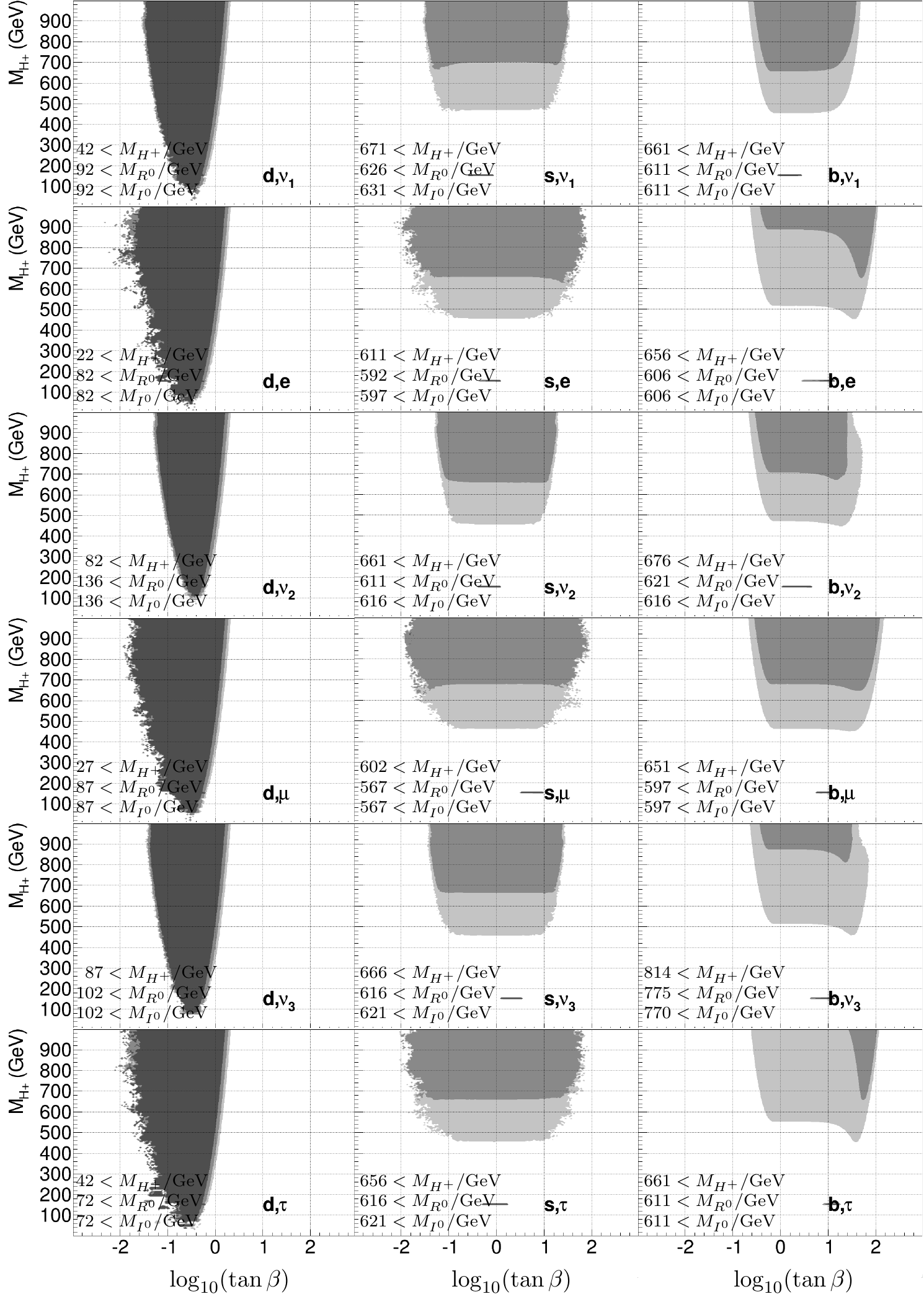}
\caption{Allowed 68\% (black), 95\% (gray) and 99\% (light gray) CL regions in $m_{H^\pm}$ vs. $\tan\beta$ for BGL models of types $(d_i,\nu_j)$ and $(u_i,\ell_j)$, i.e. for models with FCNC in the up quark sector and in the charged lepton or neutrino sector (respectively). Lower mass values corresponding to 95\% CL regions are shown in each case.\label{FIG:Results02}}
\end{figure}
\clearpage

\subsection{Acknowledgments}
The author thanks the Organisers of Discrete 2014 for the invitation to present this work for hospitality and for 
the very stimulating Conference. This work is partially supported by  Funda\c{c}\~ao para a Ci\^encia e a
Tecnologia (FCT, Portugal) through the projects CERN/FP/123580/2011,
PTDC/FIS-NUC/0548/2012,  and CFTP-FCT Unit 777
(UID/FIS/00777/2013) which are partially funded through POCTI (FEDER),
COMPETE, QREN and EU.

\end{document}